\documentclass{chjaa}            

\usepackage{graphicx}                   

\newcommand{\msun}{M$_\odot$}

\begin{document}

   \title{Supernova rates and stellar populations}

   \author{F. Mannucci \inst{}\mailto{}  }

   \institute{IRA - INAF, Largo E. Fermi 5, 50125, Firenze, Italia}
              \email{filippo@arcetri.astro.it}

\date{Received~~2001 month day; accepted~~2001~~month day}

\abstract{
We discuss the results about the nature of type Ia Supernovae 
that can be derived 
by studying their rates in different stellar populations. 
While the evolution of SN photometry and spectra can constrain the explosion 
mechanism, the SN rate depends on the progenitor system.
We review the current available data on rates as a function of parent galaxy
color, morphology, star formation rate, radio luminosity and environment.
By studying the variation of the rates with the color of the parent galaxy, 
a strong evidence was established that type Ia SNe come from both young and 
old stars. The dependence of the rates with the radio power of the parent  
galaxy is best reproduced by
a bimodal distribution of delay time between the formation of the 
progenitor and its explosion as a SN.
Cluster early-type galaxies show higher type Ia SN rate with respect to field
galaxies, and this effect can be due either to traces of young stars or to
differences in the delay time distribution.
}

\authorrunning{F. Mannucci}
\titlerunning{Type Ia SN rates and stellar populations}

\maketitle

\section{Introduction} 
\label{sect:intro}


Type Ia Supernovae are considered to be due the thermonuclear
explosion of a C/O white dwarf (WD) in a binary system due to mass accretion
from a secondary star. 
Such a conclusion follows from the a few fundamental
arguments: explosion requires a degenerate system, such as a white dwarf;
the presence of SNe Ia in old stellar systems means that at least some of
their progenitors must come from old, low-mass stars; the lack of hydrogen
in the SN spectra requires that the progenitor has lost its outer envelope;
and, the released energy
per unit mass is of the order of the energy output of the 
thermonuclear conversion of carbon or oxygen into iron. 
Considerable uncertainties about the explosion model
remain within this broad framework, such as the structure and
the composition of the exploding WD (He, C/O, or O/Ne), its mass at
explosion (at, below, or above the Chandrasekar mass) and flame 
propagation (detonation, deflagration, or a combination of the two).

Large uncertainties also remain on the nature of the progenitor binary system,
its evolution through one or more common envelope phases, and
its configuration (single or double-degenerate)
at the moment of the explosion (see \cite{yungelson05} for a review).
Solving the problem of the progenitor system is of great importance for modern
cosmology as SNe dominate metal production,
(e.g., \cite{matteucci86}), are expected to be important producer
of high-redshift dust (\cite{maiolino04a,maiolino04b,bianchi07}),
and are essential to understand the feedback process during galaxy 
formation (e.g., \cite{scannapieco06}).

While the primary observations to constrain the explosion
models are the evolution of SN photometry and spectra,
the study of the supernova rates in different types of galaxies gives strong
informations about the progenitors. For example, soon after the introduction
of the distinction between ``type I'' and ``type II'' SNe (\cite{minkowski41}),
\cite{vandenbergh59} pointed out that type IIs are frequent 
in late type galaxies ``which suggests their affiliation with Baade's
population I''. On the contrary, type Is, are 
the only type observed in elliptical galaxies and this fact ``suggests that
they occur among old stars''. This conclusion is still often accepted,
even if it is now known not to be generally valid: first, 
SN Ib/c used to be included in the broad class of ``type I'' SNe, 
and, second, it is now known that also a
significant fraction of SNe Ia have young progenitors.

\begin{figure}   
  \centerline{\includegraphics[width=7cm]{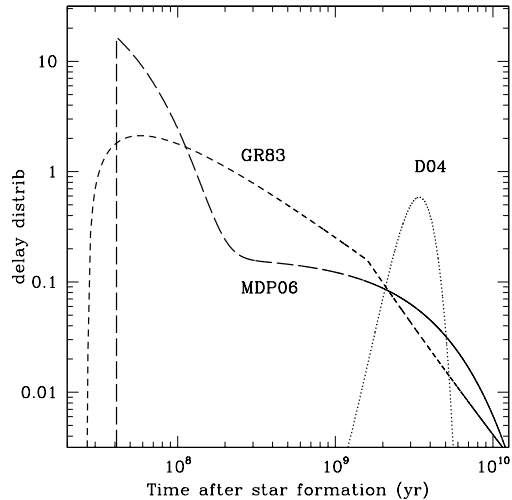}}
  \caption{
  Three examples of DTDs. 
  Short dashed: the single-degenerate model by Greggio \& Renzini (1983, GR83).
  Dotted: empirical DTD from Dahlen et al. (2004) (D04).
  Long dashed: empirical DTD similar to the ``two population'' model by
  Mannucci et al. (2006, MDP06).
  The latter is the sum of two exponentially declining
  distributions with 3 and 0.03 Gyr of decay time, respectively, each one
  containing 50\% of the events.
  }
\end{figure}

\section{The delay time distribution}

The key quantity to relate the type Ia SN rate to the parent stellar
population is the delay time distribution (DTD), i.e., the distribution 
of the delays between the formation of a binary system and its 
explosion as a SN. In general, deriving an expected DTD from a progenitor model
is not an easy task because many parameters 
are involved, such as the initial distribution of  orbital parameters in the 
binary system, the distribution of mass ratio between primary and
secondary star, the efficiency of mass loss during the common envelope phase,
the efficiency of mass transfer from one star to the other, the 
amount of mass retained by the primary star during accretion. 
Also the uncertainties in the
explosion model play a key role: for example, it is not known if it is
necessary to reach the Chandrasekar mass to start the explosion of if it is
enough to be in the sub-Chandrasekar regime.

Two basic models are present in the
literature, the single-degenerate (SD) model, in which a white dwarf
accretes hydrogen-rich mass from a non-degenerate secondary, and a
double-degenerate (DD) model, where two white dwarfs merge after 
the emission of gravitational waves.
In 1983, Greggio \& Renzini computed the expected DTD for a single degenerate
system, and such a computation was later refined and extended to the
double-degenerate systems by many authors, as
\cite{tornambe86,tornambe89,tutukov94,yungelson00,matteucci01,belczynski05}
and \cite{greggio05}. 
In many cases the results of these models are very different.
In some cases, all the explosions are concentrated in a very
narrow range of delay time (for example, in the \cite{yungelson00} SD
Chandrasekar-mass model, all the SNe explode between 0.6 and  1.5 Gyr),
in other cases the explosion happens at any delay time (from 25 Myr to 12 Gyr,
in the \cite{yungelson00} DD model); in some models all happens soon after the
formation (within 1 Gyr for the \cite{belczynski05} SD model), in other cases
the first SNe explode after a very long time (more than 10 Gyr for the 
\cite{belczynski05} semidetached double white dwarf model); some distributions
are smooth, as the  analytic models by \cite{greggio05}, some
others have multiple peaks (\cite{belczynski05}).

The observed SN rate is the convolution of DTD with the past star-formation
history (SFH) of the galaxies. This latter function also determines the stellar
population. As consequence, studying the SN rates in different parent galaxies
can put strong constraints on the DTD.

In Figure~1 we show three representative DTDs that will be discussed later. 
First, we show the DTD theoretically derived by \cite{greggio83} (GR83), which
is based on a single degenerate model; second, the empirical DTD derived 
by \cite{dahlen04} by fitting the evolution of the SN rate
with redshift; third, a simplified version of the empirical DTD 
obtained by \cite{mannucci06} 
by discussing the dependence of the rates on redshift, colors of the parent
galaxies and radio power of the hosts.

\begin{figure}   
  \centerline{\includegraphics[width=7cm]{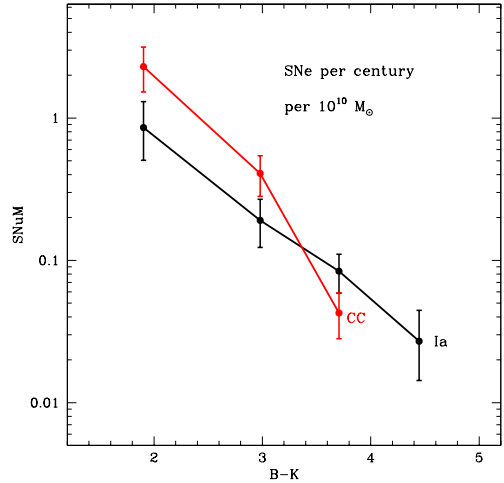}}
  \caption{SN rate per unit stellar mass as a function of the (B--K) 
  color of the
  parent galaxy (from \cite{M05}) showing the strong increase of all
  the rates toward blue galaxies}
\end{figure}

\begin{figure}   
  \centerline{\includegraphics[width=7cm]{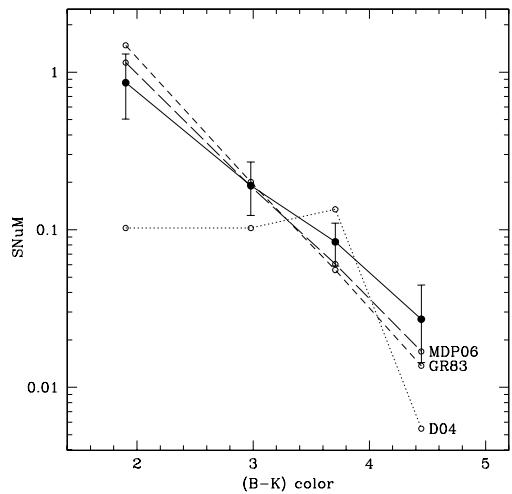}}
  \caption{Predictions of the three DTDs in Figure~1 for the
   dependence of SN rate on the (B--K) color of the
  parent galaxy. Dots with error bars are the observed rate and are connected
  by the solid line, the lines for the models are coded as in Figure~1.
  }
\end{figure}

\section{The Supernova rate per unit mass and the 
         {\em weak} bimodality in type Ia SNe}

The most important observations to constrain the DTD 
are the SN rates per unit mass in galaxies of different 
types, as presented by \cite{M05} (see Figure~2).
Using the SN rate in \cite{C99}, we found that the
bluest galaxies, hosting the highest Star Formation Rates (SFRs), 
have SN Ia rates about 30 times larger than those in the reddest, 
quiescent galaxies. 
The higher rates in actively star-forming galaxies imply that
a significant fraction of SNe must be due to young stars,
while SNe from old stellar populations are also
needed to reproduce the SN rate in quiescent galaxies. This lead
\cite{M05} to introduce the simplified 
two component model for the SN Ia rate 
(a part proportional to the stellar mass and another part to the SFR).
These results were later confirmed by \cite{sullivan06}, while
\cite{scannapieco05}, \cite{matteucci06} and \cite{calura07}
successfully applied this model
to explain the chemical evolution of galaxies and galaxy clusters.

In the DTD formalism, these results can be explained only by a DTD
spanning a wide range of delay times, with a peak at early time and
a slow decrease afterward.
This is shown in Figure~3, where the predictions of the three DTDs in Figure~1 
are shown. It is evident that the GR83 and MDP06 DTDs, 
extending from below 10$^8$ to above 10$^{10}$ years, 
reproduce the 
variation of the rates with the color of the parent galaxies. At variance, 
the D04 DTD, showing a narrow distribution peaked at about 3.5 Gyr, 
does not reproduce the global trend, 
neither the rate in the old galaxies (having ages much larger than 3.5 Gyr)
nor the overproduction of SNe in active galaxies (having a significant 
fraction of young stars).

Such a result was recently confirmed by \cite{aubourg07}. They derived the SFH
of 257 SDSS-DR5 galaxies containing SNe Ia, and demonstrated that a 
significant number of events are associated to stellar populations as 
young as $10^7$ years.

\begin{figure}   
  \includegraphics[height=6.3cm]{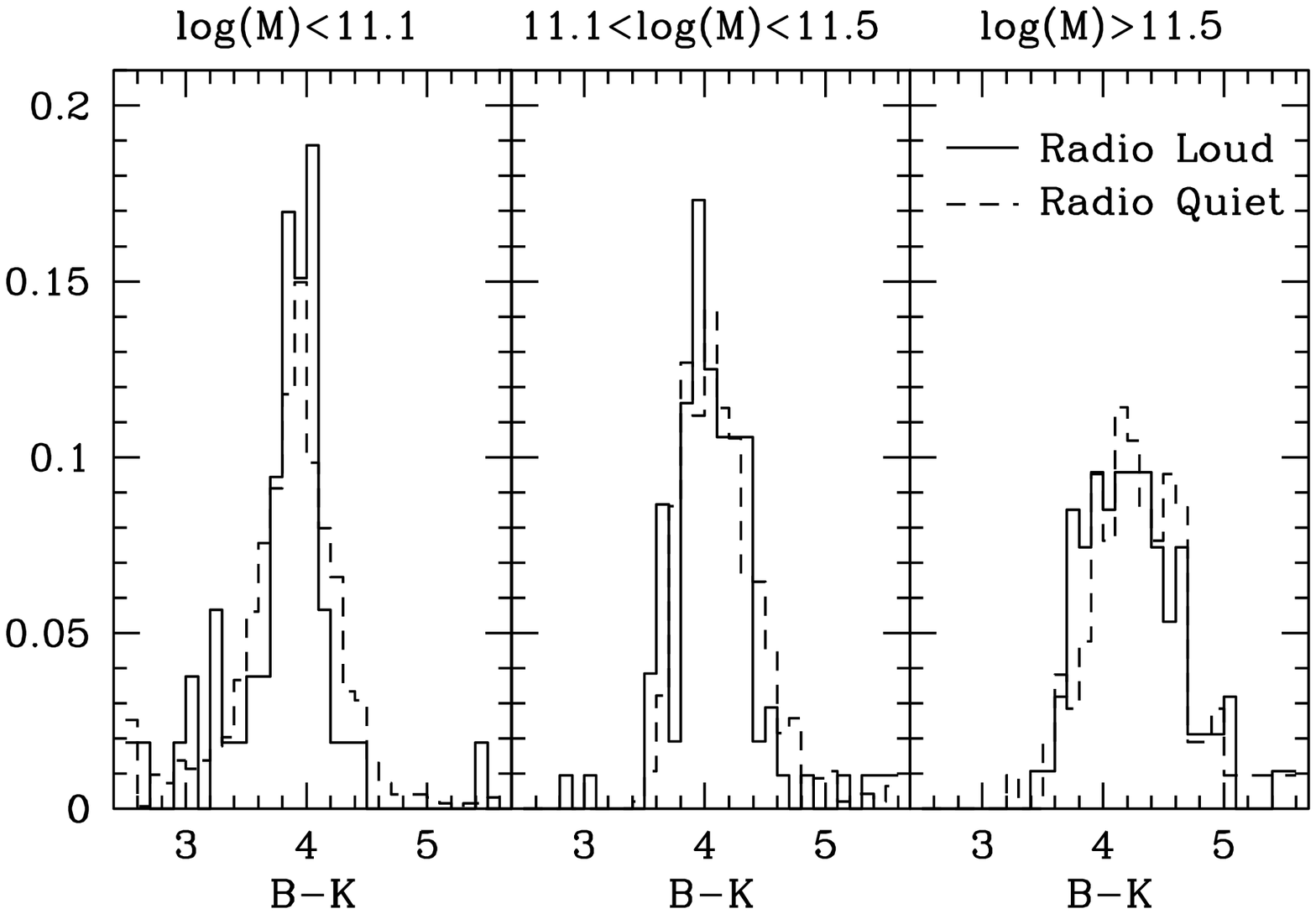}
  \includegraphics[height=6.3cm]{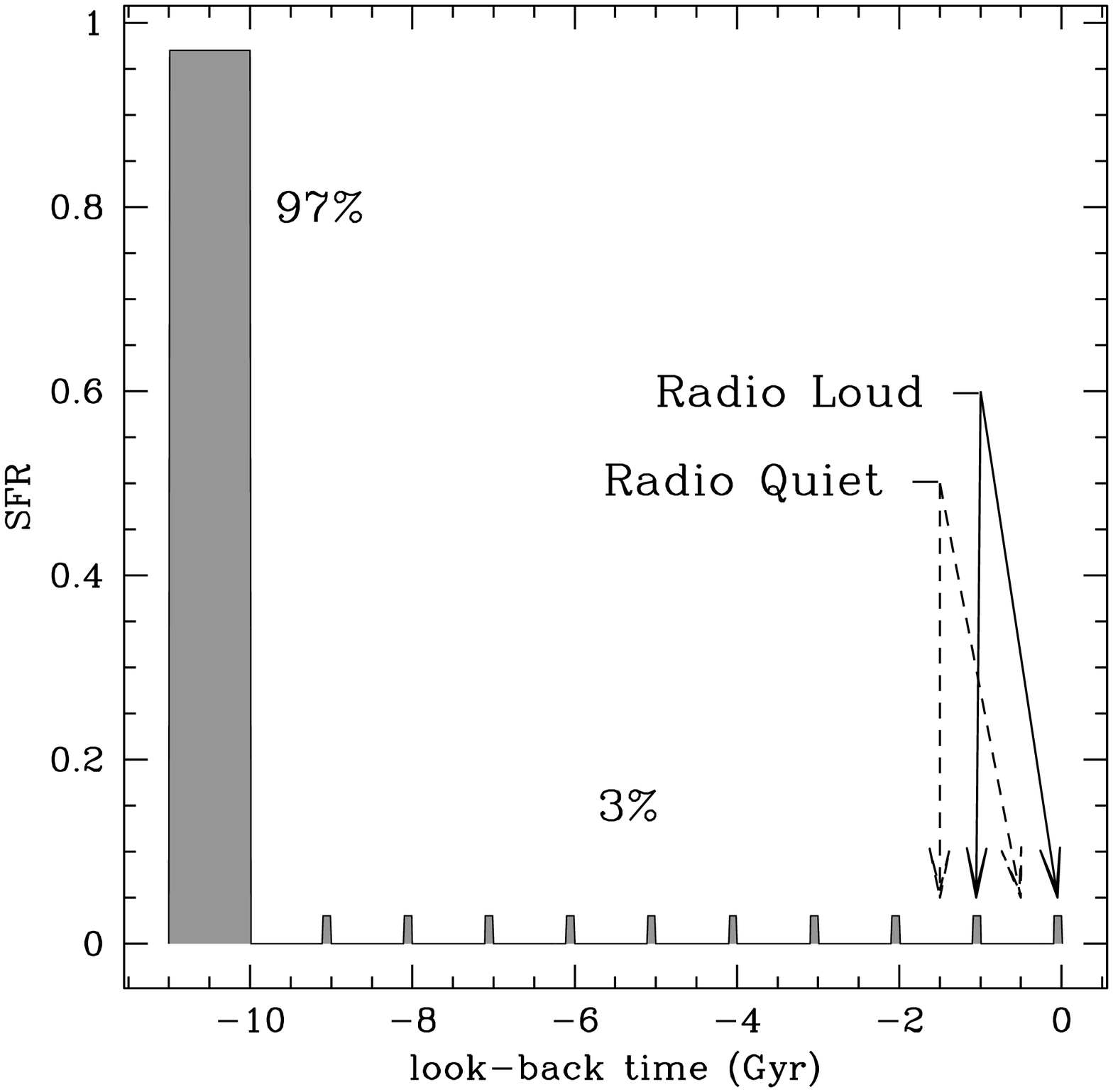}
  \caption{
  {\em Left}: (B--K) color distribution of early-type radio-loud 
  (solid line) and
  radio-quiet galaxies (dashed line) in three stellar mass ranges. The two
  groups of galaxies have practically indistinguishable color distributions, 
  meaning that the stellar populations are similar. 
  {\em Right:} Model of early-type galaxies reproducing both the dichotomy
  radio-loud/radio-faint and the similar (B--K) colors.
  }
\end{figure}

\section{The rate in radio-loud galaxies and 
         the {\em strong} bimodality in type Ia SNe}

\cite{dellavalle03} and \cite{dellavalle05} 
studied the dependence of the SN Ia rate 
in early-type galaxies on the radio power of the host.
They found that the rates in radio-loud galaxies is about a factor of 4 larger
than in radio-quiet galaxies. This is interpreted as due to
small episodes of accretion of gas or capture of small
galaxies. Such events result in both fueling the 
central black hole, producing the radio activity,
and in creating a new generation of stars, producing the increase in the
SN rate.
As the lifetime of the radio-activity is of the order of $10^8$ years,
this observation can be used to constrain the DTD at such short delay times,
once a model of galaxy stellar population is introduced.

The difference between radio-loud and radio-quiet galaxies can
be reproduced by the model of early-type galaxy shown in the right panel
of Figure~4: most of the stars are
formed in a remote past, about $10^{10}$ years ago, while a small minority
of stars are created in a number of subsequent bursts. A galaxy appears
radio-loud when is observed during the burst, radio-faint soon after,
and radio-quiet during the quiescent inter-burst period.
The abundance ratio between radio-loud and radio-quiet galaxies,
about 0.1 in our sample, means that the duty cycle
of the burst events is about 10\%. As the duration of the radio-loud phase is
about 10$^8$ years, in 10$^{10}$ years the early-type galaxies are expected to 
have experienced 10 small bursts, i.e., 
1 every 10$^9$ years.

This model naturally explains the fact that radio-loud and radio-quiet
early-type galaxies have very similar (B--K) colors,
a sensitive indicator of star formation and stellar age.
This is shown in the left panel of Figure~4, where the two
color distributions are compared.
Only a small difference in the median of the two distributions 
might be present at any mass, i.e.,
the radio-loud galaxies appear to be 0.03-0.06 mag bluer than radio-quiet
galaxies, 
and this could be the effect of last on-going burst of star formation.

The amount of mass in younger stars 
can be estimated from the (B--K) color, that is  consistent
with the value of (B--K)$\sim$4.1 typical of old stellar populations.
By using the \cite{bruzual03} model, we obtain that no more than
3\% of stellar mass in total can be created in the 10 bursts 
(0.3\% of mass each) if we assume negligible extinction, 
otherwise the predicted color would be too blue.
The maximum mass in new stars can reach 5\% 
assuming an average extinction of the new component of $A_V=1$. 

This model predicts that traces of small amounts of recent star formation 
should be present in most of the local early-type galaxies. This is actually
the case: most ellipticals show very faint emission lines
(\cite{sarzi06,haines07}),
tidal tails (\cite{vandokkum05}), dust lanes (\cite{colbert01}),
HI gas (\cite{morganti06}), molecular gas (\cite{welch03}), and
very blue UV colors (\cite{schawinski07}).
Furthermore, \cite{ferreras06} have found
evidence for recent star formation, at the percent level, in ellipticals
in compact groups, but not in field ellipticals.
Even if the interpretation of most of these effects is matter of debate
(for example, \cite{haines07} have found higher levels of 
present star formation in field rather than cluster early-type galaxies),
they could be at least partially related to the higher
SN Ia rate in cluster early-type galaxies and in radio-loud objects.

\begin{figure}   
  \centerline{\includegraphics[width=7cm]{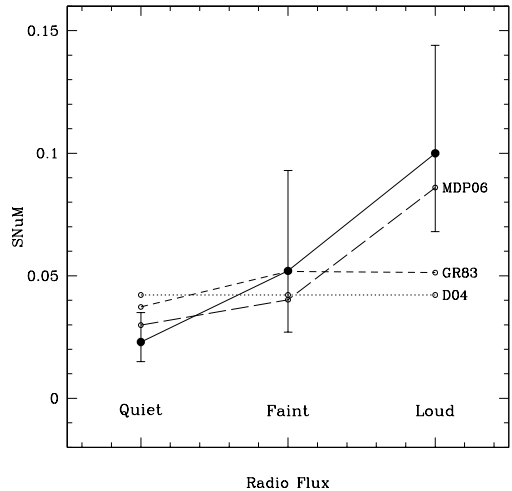}}
  \caption{
  The solid dots with error bars show the type Ia SN rate as a 
  function of the radio power of the parent galaxy. The other lines,
  coded as in Figure~1, show the results of the GR83, D04 and MDP06 models
  }
\end{figure}

Using this model with a total fraction of new stars of 3\%, we derive the
results shown in figure~5. 
We see that the theoretical models by \cite{greggio83}
predicts too few SNe in the first
$10^8$ years (about 11\%) to accurately fit figure~5.
Also the D04 DTD is not adequate.
The observed rates can be reproduced only by
adding a ``prompt'' component (in this case modeled in terms 
of an exponentially
declining distribution with $\tau=$0.03~Gyr) to a ``tardy'' component 
(an other declining exponential with $\tau=$3~Gyr), 
each one comprising 50\% of the total number of events.

It should be noted that this {\em strong} bimodality is based on a small number
of SNe (21) in early-type galaxies, and the results of oncoming larger SN
searches are needed to  confirm (or discard) this result.

\begin{figure}   

\centerline{\includegraphics[height=11cm,angle=270]{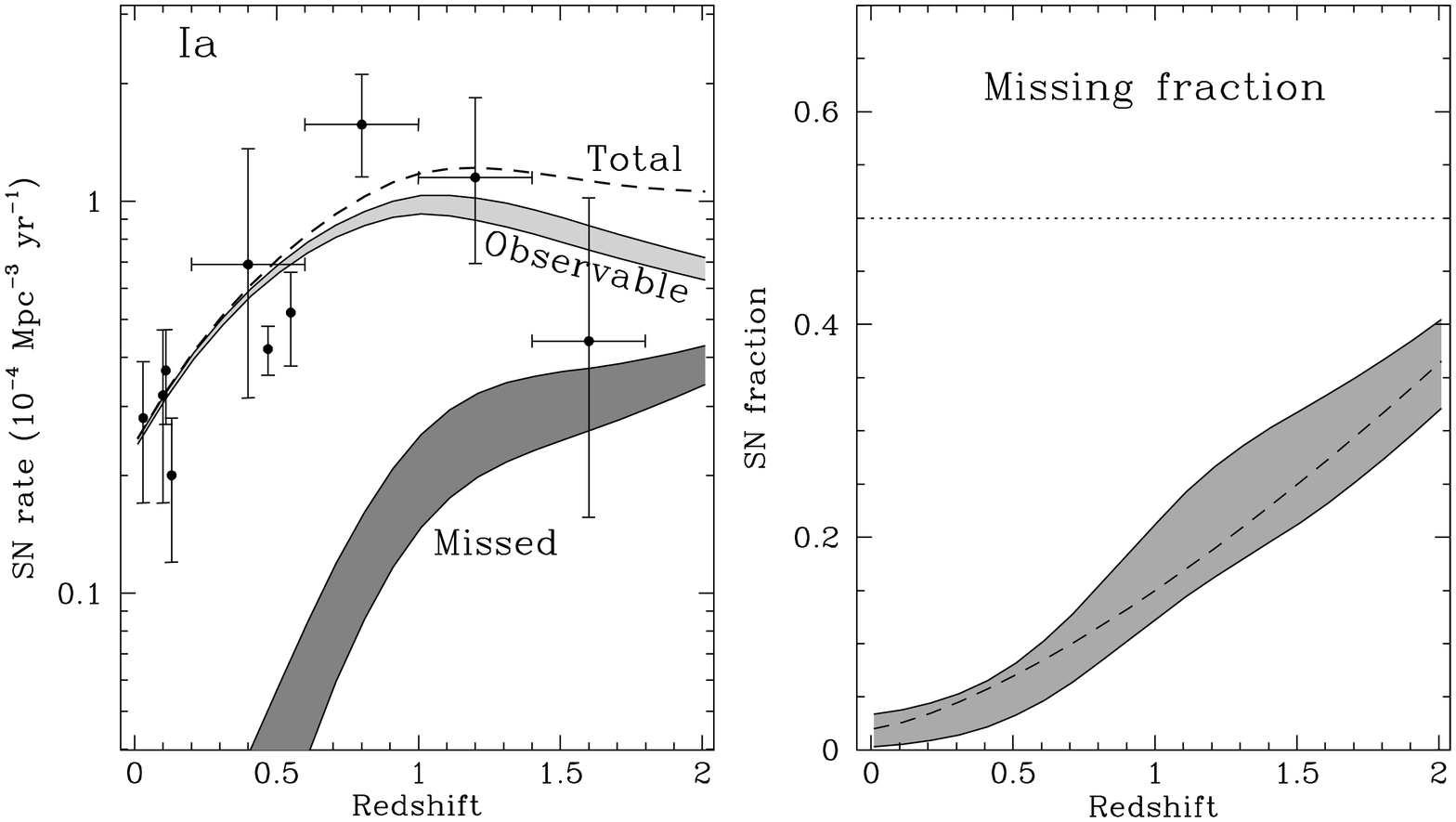}}
\centerline{\includegraphics[height=11cm,angle=270]{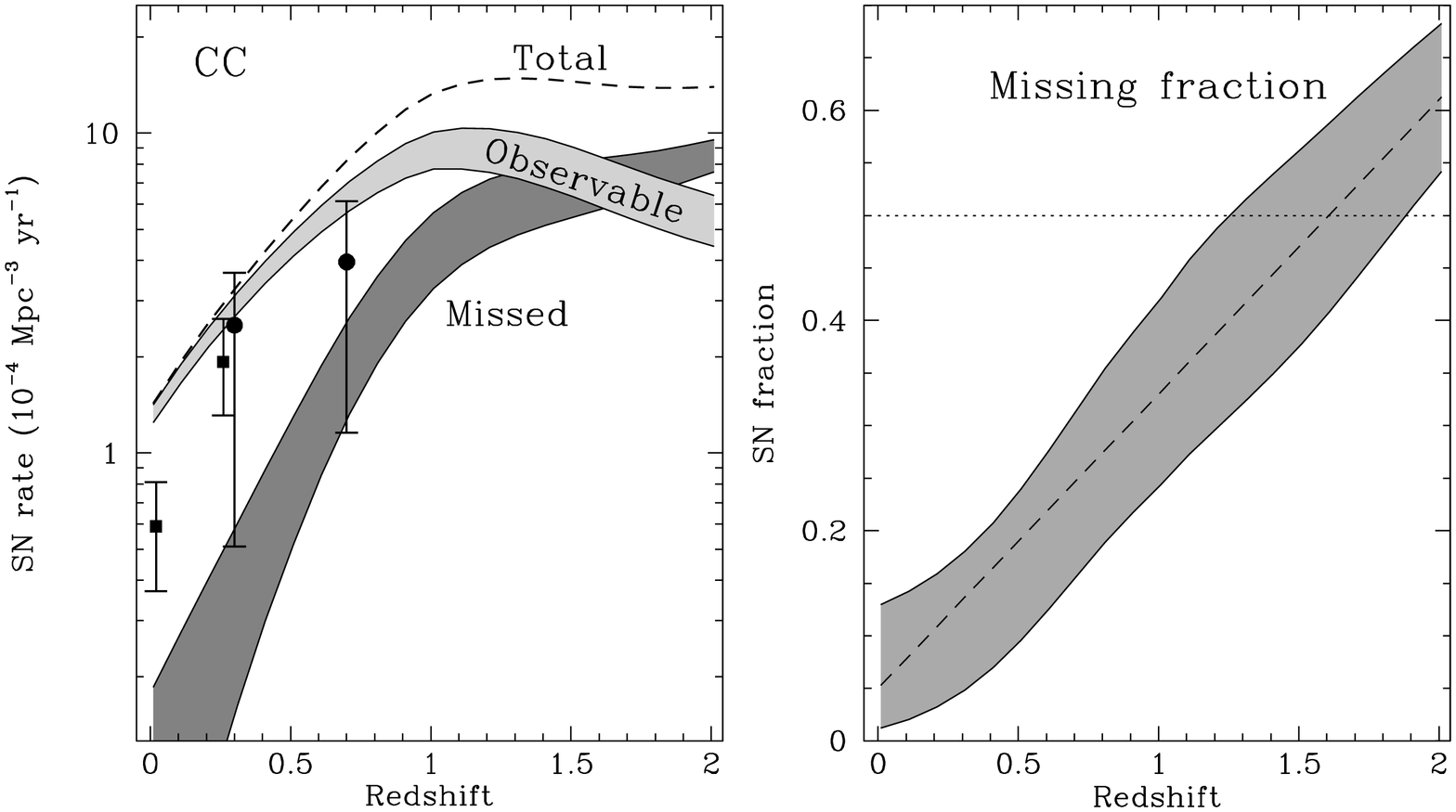}}

  \caption{
  Evolution of the rates of type Ia (top panels) 
  and core-collapse SNe (bottom panels), from
  \cite{mannucci07a}. In the left panels,
  the dashed line shows the total rate
  expected from the cosmic star formation history, the light grey area
  the rate of SNe that can be recovered by the optical and near-IR searches,
  and the dark grey area the rate of SNe exploding inside dusty
  starbursts and which will be missed by the searches.
  The right panels show the fraction of missed SNe.
  }
\end{figure}

\section{Evolution of the SN rate with redshift}

Constrains on the DTD can also be derived by comparing the redshift
evolution of the SN 
rate with the cosmic SFH. 
This possibility was explored by many authors, as
\cite{madau98,dahlen99,pain02,dahlen04,M05,barris06,neill06,botticella07}
and \cite{poznanski07}.
Unfortunately, large uncertainties are present on both functions:
(1) SN rates published by different groups often differ much more than 
the quoted errors, implying the presence of uncorrected systematics;
(2) many different SFHs can be fitted to the present data (see
\cite{hopkins06} and \cite{mannucci07a} for reviews of the available data).
As a consequence, \cite{forster06} demonstrated that no strong conclusions 
on the DTD can now be reached by only using the cosmic 
evolution of the rate.

An additional complication is the amount of dust affecting both the SN searches
and the determination of the SFH. 
Near-infrared and radio searches for core-collapse supernovae
in the local universe 
(\cite{maiolino02,mannucci03,lonsdale06,mattila07,cresci07})
have shown
that the vast majority of the events occurring in massive starbursts
are missed by current optical searches because they explode in
very dusty environments. 
Recent mid- and far-infrared observations (see \cite{perez05}
and references therein) have shown that the fraction of 
star-formation activity that takes place in very luminous 
dusty starbursts sharply increases with redshift and becomes the dominant 
star formation component at z$\ge$0.5. 
As a consequence, an increasing fraction of SNe are expected to
be missed by high-redshift optical searches. 
By making reasonable assumptions on the number of SNe that can be observed by
optical and near-infrared searches in the different types of galaxies
(see \cite{mannucci07a} for details)
we obtain the results shown in figure~6. We estimate that 
5--10\% of the local core-collapse (CC) SNe are out of reach of the
optical searches.
The fraction of missing events rises sharply toward z=1, where about 
30\% of the CC SNe will be undetected. At z=2 the missing 
fraction will be about 60\%. 
Correspondingly, for type Ia SNe, our computations provide 
missing fractions of 15\% at z=1 and 35\% at z=2.
Such large corrections are crucially important to compare
the observed SN rate with the expectations from the
evolution of the cosmic star formation history, 
and to design the future SN searches at high redshifts.

\section{SN rate in cluster and field galaxies}

One effect preventing a good determination of the DTD is the complex SFH
of most galaxies in the local universe. 
A possible solution to this problem is to measure
the SN rate in galaxy clusters. In these systems,
most of the stellar mass is contained in elliptical 
galaxies, whose stellar populations are dominated by old stars (see the
discussion in section~3). 
In principle, this significant reduction in the uncertainty in the SFH
is of great help in deriving the DTD.

The cluster SN rate is also of great importance to study the
metallicity evolution of the universe.
The gravitational potential well of galaxy clusters is deep enough to retain
in the intracluster medium (ICM) all the metals which are produced in 
galactic or intergalactic SNe.
As a result, the metallicity of the ICM
is a good measure of the integrated past history
of cluster star formation and metal production. 
As discussed by \cite{renzini93}, the measured 
amount of iron is an order of magnitude too high to be produced 
by the SN Ia exploding at the current rate. 
The proposed explanations of this effect include the presence of higher SN
rates in the past (\cite{matteucci06}), the contribution of the intracluster
stellar population (\cite{zaritsky04}), and the existence of
evolving properties of
star formation processes (\cite{maoz04,lowenstein06}).
\cite{calura07} used the observed cosmic evolution
of iron abundances to constrain the history
of SN explosion, iron formation and gas stripping in galaxy clusters. 
They found a good
agreement with the observations, especially when the ``two channel'' model
of SNe Ia by \cite{M05} is used.

There are strong motivations for measuring the cluster rates of the other
physical class of SNe, the core-collapse group.
The CC SNe rate per unit mass
is sensitive to the IMF, because
the SN explosions are due to massive stars while most of the mass
is locked in low-mass stars. As a consequence, the study the CC SN rate as a
function of environment is a sensitive test for any systematic
difference in IMF.

\smallskip

Prompted by all these reason, several groups have searched for 
SNe in galaxy clusters 
(\cite{crane77,barbon78,norgaard89,galyam02,galyam03,germany04,sharon07})
but the results are still quite sparse and based on a small number of SNe.
For this reason 
we used the SN sample described by \cite{C99}, comprising 136 SNe,
to measure the SN rate in galaxy clusters, as described in \cite{mannucci07b}.

A galaxy is considered to be part
of a cluster if its projected distance from a known cluster is below 
1.5 Mpc, and if its velocity difference in below 1000 Km/sec
(\cite{dressler97,hansen05}).
Of the 8349 galaxies of the full sample, 1666 (about 20\%) 
belongs to clusters.
The expected strong morphological segregation is well recovered,
with cluster having a
a much larger fraction of early-type galaxies (53\%) than the field (19\%).
Cluster galaxies host 44 SNe (32\% of the total), and field galaxies the
remaining 92. 

The above classification
has a number of weaknesses. First, we assume that all clusters 
have the same radial extent, even if this is known not to be true.
Second, clusters show a galaxy density that smoothly decreases with radius 
rather than a sharp cutoff.
Third, the cluster
catalog is not complete and it is possible that some clusters are missing.
All these effects are likely to produce some degree of misclassification
in both directions. However, any missclassification 
can dilute or hide an existing difference in rates,
but it is unlikely to produce an artificial difference in rates
or enlarge a small difference.

Figure~7 shows the resulting SN rate per unit mass as a function of 
galaxy morphology for cluster and field galaxies, for both CC and Ia SNe.

\begin{figure*}		
\centerline{\includegraphics[angle=-90,width=15cm]{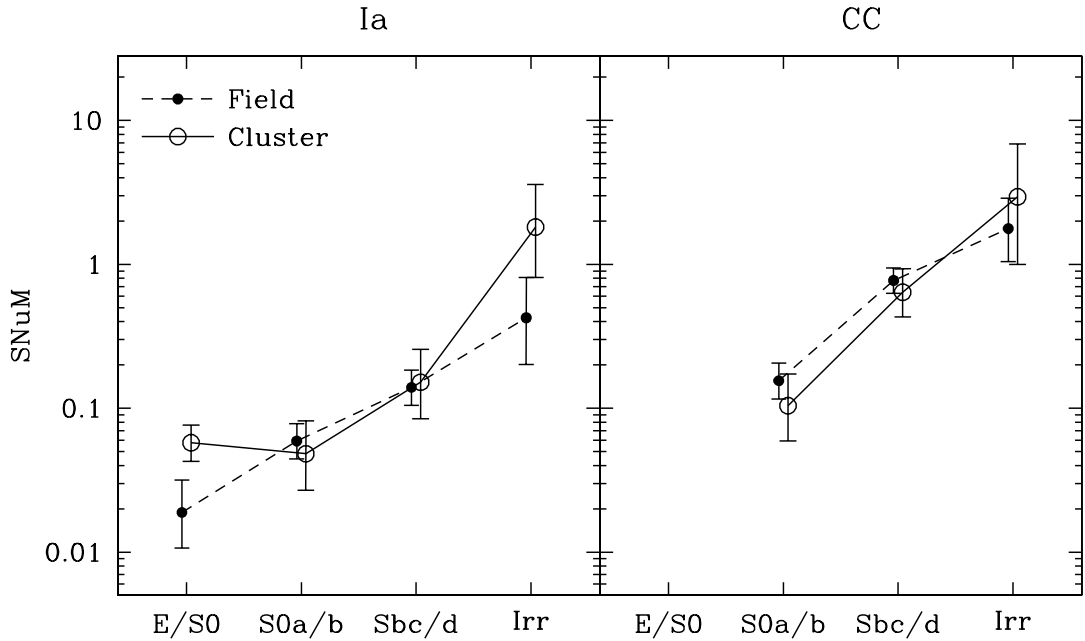}}
\caption{
\label{fig:massrate}
SN rates per unit mass as a function of galaxy morphology 
for type Ia ({\em left}) and CC SNe ({\em right})
in clusters (empty dots) and in the field (solid dots). 
}
\end{figure*}

\subsection{Core-collapse supernovae}
\label{sec:ccrate}

About 36\% of the cluster SNe (16 out of 44) are CC, and this allows for 
the first measurement of the CC rate in galaxy cluster.
As shown in the panel b 
of Figure.~\ref{fig:massrate}, they are hosted by the cluster
spirals and irregular galaxies, and we do not find any significant
difference in CC SN rate between cluster and field galaxies. 
As a consequence, we don't detect any difference in the IMF in the two 
environments.
It should be noted that two effects 
are present that could hide any intrinsic difference: 1) our determination
of galaxy membership is not perfect, as discussed above,  
and 2) in clusters, 
the number of CC SNe per galaxy type is small,
and only large differences of the order of 50\% or more
could be significantly detected.\\

\subsection{Type Ia supernovae}
\label{sec:iarate}

Figure~7 shows that the sharp increase of the rate from early-type
to late-type galaxies and irregulars is present both in clusters and in the 
field. All the rates are comparable in the two environments, 
with the exception of early-type galaxies.

Clusters, comprising 20\%
of all the early-type galaxies and 49\% of the total ``sensitivity'' (i.e.,
the product of stellar mass times control time) for these galaxies, 
contains 15 out of the 20 SN Ia events in early-type galaxies,
i.e., 75\% of the total sample. As a consequence, they show a higher rate
($0.058^{+0.019}_{-0.015}$ SNuM) 
than field early-type galaxies ($0.019^{+0.013}_{-0.008}$ SNuM). 

\smallskip

The statistical significance of the rate difference can be estimated in several
ways. The most straightforward is to apply the $\chi^2$ test by 
considering that numbers of SNe are effected by Poisson
errors. 
The null hypothesis, i.e., the hypothesis of no rate
difference between cluster and field, would predict 9.9 SNe (instead of 15)
in clusters and 10.1 (instead of 5) in the field.
By applying the $\chi^2$ for 1 degree-of-freedom 
we obtain that the statistical significance of the difference is 97.5\%.

Another way to compute the statistical significance of the difference
is to consider the binomial distribution of the probability
of a SN in an early-type galaxy to explode in the cluster or in the
field. The probability of detecting 15 SNe or more 
in clusters having an intrinsic probability of 49\% is  1.9\%, i.e., 
we can exclude that such a number comes from a random distribution with a
confidence level higher than 98.1\%, in good agreement with the previous method.

As a consequence the difference is statistically significant but not at
a level above any doubt. Also, the existence
of possible low-level systematics (such as non complete corrections
for the differences among the different searches in the \cite{C99} sample)
means that a larger number of SNe 
in a more homogeneous sample of galaxies are needed to
confirm or exclude such an effect.

We note that the cluster SN rate do not take into account the
possible contribution from intergalactic SNe, which \cite{galyam03}
quantified in about 20\% of the total. Such an extra rate should be added to
the cluster rate and not to the field one, increasing the measured difference
and its statistical significance.

\smallskip

In principle, 
the dependence of the rate with environment 
could be just an indirect effect due to the different distribution of cluster 
and field galaxies according to some other property of the galaxies
more directly affecting the rates. In the following, we
analyze three possibly important properties: 
stellar mass, morphology and radio activity.

Galaxy ``downsizing'', i.e. the observation that less massive systems form 
their stars after and more slowly than more massive systems, 
has recently become the standard paradigm to describe galaxy
formation. 
The dependence of the evolutionary paths of the galaxies on stellar mass
rises the possibility that specific SN Ia rate in early-type galaxies
also depends on this parameter.
In our case, mass does not appear to be the main
driver of the rate difference, for two reasons: 1) 
cluster and field galaxies have similar mass distributions,
with differences limited to below 10\%;
2) the distribution of the number of detected SNe with galaxy mass 
follows closely that of the ``sensitivity'', i.e., galaxies of different mass
have the same SN rate per unit mass.

At variance, both radio-power and morphology appear to affect the rate, 
with radio-loud and S0 galaxies being more prolific than radio-quiet and
elliptical galaxies, respectively. 
Nevertheless, in both cases the environment
plays a role: both E and S0 are more active in clusters, 
and both radio-loud and radio-quiet galaxies have higher rates 
in clusters than in the field. As a consequence the environments is
another parameter that independently affects the rates.


\smallskip

The interpretation of this possible difference in type Ia SN rate
between cluster and field early-type galaxies is not straightforward.
As the observed rate is the convolution of the SFH with the DTD, the
differences could be due to either of these functions.

\cite{M05}, \cite{sullivan06}, and \cite{aubourg07}
have shown that SN Ia rate has a strong dependence on the 
parent stellar
population, with younger stars producing more SNe. 
The difference in SN rate could be related to this effect, i.e., 
to a higher level of recent star formation in cluster ellipticals.
Only a very small amount of younger stars are needed, because
the amplitude of the DTD 
at short times can be hundreds of times larger than at
long times.
As an example, the GR83 single-degenerate model
has 300 more power at $10^8$ yr than at $10^{10}$ yr, and this means
that a recent stellar population of 0.3\% in mass can provide
as many SNe as the remaining 99.7\% of old stars. 
The MDP06 ``two channel'' model this ratio can be as large as a factor of
1000. 

The presence of small amounts of young stars in early-type galaxies, 
and its larger presence in cluster galaxies, is not in contrast with
observations. Actually, many ellipticals shows signs of recent 
interactions or activity, as discussed in section 4.

If this is the correct interpretation, the ``prompt'' population of 
SNe Ia would be associated to the explosion of CC SNe from the same young
stellar populations.
If a SN Ia is to explode within $10^8$ yr from the formation of its progenitor, 
the primary star of the progenitor binary system 
must have a mass above 5.5 \msun\ to allow for the 
formation of a white dwarf in such a short time. 
\cite{mannucci06} have shown that the observed rates of type Ia SNe
implied that about 7\% of all stars between 5.5 and 8 \msun\ 
explode as ``prompt'' SNe Ia, while the ``tardy'' population corresponds to 
a lower explosion efficiency, about 2\%, and on a much longer timescale.
For a Salpeter IMF and assuming that all the stars between 8 and 40 \msun\ 
end up at CC SNe, we expect 1.3 CC SNe for each ``prompt'' type Ia.
Assuming that the difference between cluster and field early-type galaxies is
due to the ``prompt'' SNe Ia, the rate of this population is of the order of 
0.066-0.019=0.047 SNuM. 
Converting this rate in observed number, 
about 2 CC SNe are expected in the cluster early-type galaxies
of our sample, consistent with our null detection at about 1.3$\sigma$
level.
We conclude that the non detection of CC in the early-type galaxies
belonging to our sample is
consistent with the hypothesis of a ``prompt'' Ia component.
We also note that a few well-classified CC SNe have been discovered in the
recent past in prototypical Ellipticals, such as the type Ib SN 2000ds 
in NGC2768.

\smallskip

A second possible interpretation is that the higher rate in cluster 
early-type galaxies is related  
to differences in the DTD. If the stars in ellipticals 
are 9-12 Gyrs old (see, for example, \cite{mannucci01}), the SN rate
is dominated by the tail of the DTD at long times. 
Differences in the environments could produce small differences in the
shape of this function. For example, 
during a galaxy-galaxy encounter or merging, the small number of close 
gravitational interaction between a binary systems
and a star of the other galaxy could sometime produce the shrinking of 
the separation of old binary systems. With simple order-of-magnitude 
considerations we estimate that such an effect is likely to give a small,
but possibly non-negligible, contribution.
Only a few percent of the binary systems give rise
to a SN during a Hubble time, and only a small fraction of them explode
more then 8 Gyr after formation. Therefore such a mechanism would be 
effective even if is it is able to shrink a small fraction of binary systems,
of the order of 0.5\%. 
It is also possible
that new binary systems are created during the interaction, 
or that isolated white dwarf accreted gas of the other galaxy. These
effects could be more efficient in clusters, where the volume density 
of the galaxies is larger.

An interesting possibility is also that the changes in the DTD are 
related to differences in metallicity
between cluster and field early-type galaxies, as discussed by
\cite{sanchez06,bernardi06,collobert06} and \cite{prieto07}. 
The differences between cluster and field galaxies presented by these
authors are neither large nor always in the same direction. Nevertheless
systematic, although not large, differences in metallicity could be present 
and produce significant changes in the
DTD, for example, by affecting the efficiency of mass loss during
the complex life of a binary system.

\section{Summary and conclusions}

We have reviewed the deep relations between SN rate and stellar populations. 
All the observations available for CC SNe confirm that these SNe derive from 
young massive stars. Nevertheless, the use of the CC SN rate as a measure
of star formation requires a good knowledge of dust extinction which is expected to vary rapidly with cosmic time.
Current observations of the SN rates in different classes of galaxies
show that SNe Ia derive from both young and old stellar systems, and 
between 10\% (weak bimodality) and 50\% (strong bimodality) of the events must 
explode within $10^8$ yrs after formation.
Whether this is due to the presence of two different channels of explosion
(such as single- and double-degenerate) or to a bimodal distribution of
some parameter of the exploding systems (such as the mass ratio between the two
stars of the binary system) is currently not known.


\begin{acknowledgements}
Most of the results described here were obtained in collaboration with
N. Panagia, M. Della Valle, D. Maoz, K. Sharon, M. T. Botticella, and
A. Gal-Yam, to whom I am deeply indebted. 
\end{acknowledgements}


\end{document}